# Two-stage Plant Species Recognition by Combining Local K-NN and Weighted Sparse Representation

Shanwen Zhang[1], Harry Wang [2,3*], Wenzhun Huang[1]

1. Xijing University, Xi'an 710123, China
2. GoPerception Laboratory, Ithaca 14850, NY, USA
3. Cornell University, Ithaca 14853, NY, USA

\* hw496@goperception.com

**Abstract**: In classical sparse representation based classification (SRC) and weighted SRC (WSRC) algorithms, the test samples are sparely represented by all training samples. They emphasize the sparsity of the coding coefficients but without considering the local structure of the input data. Although the more training samples, the better the sparse representation, it is time consuming to find a global sparse representation for the test sample on the large-scale database. To overcome the shortcoming, aiming at the difficult problem of plant leaf recognition on the large-scale database, a two-stage local similarity based classification learning (LSCL) method is proposed by combining local mean-based classification (LMC) method and local WSRC (LWSRC). In the first stage, LMC is applied to coarsely classifying the test sample. $k$ nearest neighbors of the test sample, as a neighbor subset, is selected from each training class, then the local geometric center of each class is calculated. $S$ candidate neighbor subsets of the test sample are determined with the first $S$ smallest distances between the test sample and each local geometric center. In the second stage, LWSRC is proposed to approximately represent the test sample through a linear weighted sum of all $k \times S$ samples of the $S$ candidate neighbor subsets. The rationale of the proposed method is as follows: (1) the first stage aims to eliminate the training samples that are ''far'' from the test sample and assume that these samples have no effects on the ultimate classification decision, then select the candidate neighbor subsets of the test sample. Thus the classification problem becomes simple with fewer subsets; (2) the second stage pays more attention to those training samples of the candidate neighbor subsets in weighted representing the test sample. This is helpful to accurately represent the test sample. Experimental results on the leaf image database demonstrate that the proposed method not only has a high accuracy and low time cost, but also can be clearly interpreted.

# 1. Introduction

Similarity-based-classification learning (SCL) methods make use of the pair-wise similarities or dissimilarities between a test sample and each training sample to design the classification problem. K-nearest neighbor (K-NN) is a non-parametric, simple, attractive, relatively mature pattern SCL method, and is easy to be quickly achieved [1,2]. It has been widely applied to many applications, including computer vision, pattern recognition and machine learning [3,4]. Its basic processes are: calculating the distance (as dissimilarity or similarity) between the test sample $y$ and each training sample, selecting $k$ samples with $k$ minimum distances as the nearest $k$ neighbors of $y$, finally determining the category of $y$ that most of the nearest $k$ neighbors belong to. In weighted K-NN, it is useful to assign weight to the contributions of the neighbors, so that the nearer neighbors contribute more to the classification method than the more dissimilarity ones. One of the disadvantages of K-NN is that, when the distribution of the training set is uneven, K-NN may cause misjudgment, because K-NN only cares the order of the first $k$ nearest neighbor samples but does not consider the sample density. Moreover, the performance of K-NN is seriously influenced by the existing outliers and noise samples. To overcome these problems, a number of local SCL (LSCL) methods have been proposed recently. The local mean-based nonparametric classifier (LMC) is said to be an improved K-NN, which can resist the noise influences and classify the unbalanced data [5,6]. Its main idea is to calculate the local mean-based vector of each class as the nearest $k$ neighbor of the test sample, and the test sample can be classified into the category that the nearest local mean-based vector belongs to. One disadvantage of LMC is that it cannot well represent the similarity between multidimensional vectors. To improve the performance of LMC, Mitani et al. [5] proposed a reliable local mean-based K-NN algorithm (LMKNN), which employs the local mean vector of each class to classify the test sample. LMKNN has been already successfully applied to the group-based classification, discriminant analysis and distance metric learning. Zhang et al. [6] further improved the performance of LMC by utilizing the cosine distance instead of Euclidean distance to select the $k$ nearest neighbors. It is proved to be better suitable for the classification of multidimensional data.

Above SCL, LMC and LSCL algorithms are often not effective when the data patterns of different classes overlap in the regions in feature space. Recently, sparse representation based classification (SRC) [8], a SCL modified manner, has attracted much attention in various areas. It can achieve better classification performance than other typical clustering and classification methods such as SCL, LSCL, linear discriminant analysis (LDA) and principal component analysis (PCA) [7] in some cases. In SRC [9], a test image is encoded over the original training set with sparse constraint imposed on the encoding vector. The training set acts as a dictionary to linearly represent the test samples. SRC emphasizes the sparsity of the coding coefficients but without considering the local structure of the input data [10,11]. However, the local structure of the data is proven to be important for the classification tasks. To make use of the local structure of the data, some weighted SRC (WSRC) and local SCR (LSRC) algorithms have been proposed. Guo et al. [12] proposed a similarity WSRC algorithm, in which, the similarity matrix between the test samples and the training samples can be constructed by various distance or similarity measurements. Lu et al. [13] proposed a WSRC algorithm to represent the test sample by exploiting the weighted training samples based on $l_1$-norm. Li et al. [14] proposed a LSRC algorithm to perform the sparse decomposition in local neighborhood. In LSRC, instead of solving the $l_1$-norm constrained least square problem for all of training samples, they solved a similar problem in the local neighborhood of each test sample.

SRC, WSRC, similarity WSRC and LSRC have something in common, such as, the individual sparsity and local similarity between the test sample and the training samples are considered to ensure that the neighbor coding vectors are similar to each other if they have strong correlation, and the weighted matrix is constructed by incorporating the similarity information, the similarity weighted $l_1$-norm minimization problem is constructed and solved, and the obtained coding coefficients tend to be local and robust.

Leaf based plant species recognition is one of the most important branches in pattern recognition and artificial intelligence [15-18]. It is useful for agricultural producers, botanists, industrialists, food engineers and physicians, but it is a NP-hard problem and a challenging research [19-21], because plant leaves are quite irregular, it is difficult to accurately describe their shapes compared with the industrial work pieces, and some between-species leaves are different from each other, as shown in Fig1.A and B, while within-species leaves are similar to each other, as shown in Fig.1C [22].

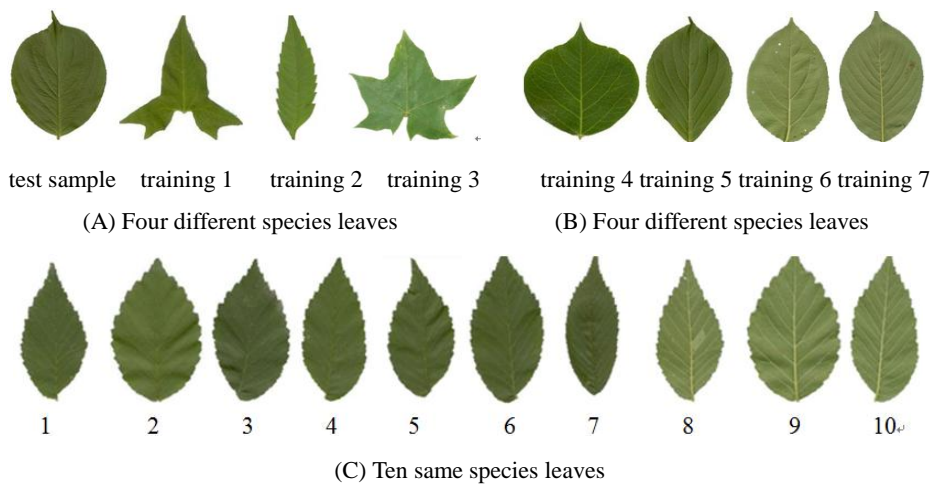

test sample   training 1   training 2   training 3       training 4 training 5 training 6 training 7
(A) Four different species leaves                (B) Four different species leaves

1   2   3   4   5   6   7   8   9   10
(C) Ten same species leaves
Fig.1 plant leaf examples

SRC can be applied to leaf based plant species recognition [23,24]. In theory, in SRC and modified SRC, it is well to sparsely represent the test sample by too many training samples. In practice, however, it is time consuming to find a global sparse representation on the large-scale leaf image database, because leaf images are quite complex than face images. To overcome this problem, in the paper, motivated by the recent progress and success in LMC [6], modified SRC [12-14], two-stage SR [25] and SR based coarse-to-fine face recognition [26], by creatively integrating LMC and WSRC into the leaf classification, a novel plant recognition method is proposed and verified on the large-scale dataset. Different from the classical plant classification methods and the modified SRC algorithms, in the proposed method, the plant species recognition is implemented through a coarse recognition process and a fine recognition process.

The major contributions of the proposed method are (1) a two-stage plant species recognition method, for the first time, is proposed; (2) a local WSRC algorithm is proposed to sparsely represent the test sample; (3) the experimental results indicate that the proposed method is very competitive in plant species recognition on large-scale database.

The remainder of this paper is arranged as follows: in Section 2, we briefly review LMC, SRC and WSRC. In Section 3, we describe the proposed method and provide some rationale and interpretation. Section 4 presents experimental results. Section 5 offers conclusion and future work.

## 2. Related works

In this section, some related works are introduced. Suppose $n$ training samples, $x_1, x_2, ..., x_n$, from $C$ different classes $\{X_1, X_2, ..., X_C\}$. $n_i$ is the sample number of the $i^{th}$ class, then $n_1 + n_2 + ... + n_C = n$.

### 2.1 LMC

Local mean-based nonparametric classification (LMC) is an improved K-NN method [6]. It uses Euclidean distance or cosine distance to select nearest neighbors and measure the similarity between the test sample and its neighbors. In general, the cosine distance is more suitable to describe the similarity of the multi-dimensional data.

LMC is described as follows, for each test sample $y$,

Step 1: Select $k$ nearest neighbors of $y$ from the $j$th class $X_j (j = 1, 2, ..., C)$, as a neighbor subset represented by $N_k(y, X_j)$;

Step 2: Calculate the local mean-based vector for each class $X_j$ by $N_k(y, X_j)$,

$$\overline{x_j} = \frac{1}{k} \sum_{j=1}^{k} x_j, x_j \in N_k(y, X_j) \qquad (1)$$

Step 3: Calculate the distance $d(y, \overline{x_j})$ between $y$ and $\overline{x_j}$.

Step 4: $\text{classify}(y) = \arg\min_j d(y, \overline{x_j})$ if Euclidean distance metric is adopted; while $\text{classify}(y) = \arg\max_j d(y, \overline{x_j})$ if cosine distance metric is adopted.

## 2.2 SRC

SRC relies on a distance metric to penalize the dissimilar samples and award the similar samples. Its main idea is to sparsely represent and classify the test sample by a linear combination of all the training samples. The test sample is assigned into the class that produces the minimum residue.

SRC is described as follows,

Input: $n$ training samples $x_i \in R^m (i=1,2,...,n, m<n)$, a test sample $y \in R^m$.

Output: the class label of $y$.

Step 1: Construct the dictionary matrix $A \in R^{m \times n}$ by $n$ training samples $x_i(i=1,2,...,n)$. Each column of $A$ is a training sample called basis vector or atom. Normalize each column of $A$ to unit $l_2$-norm.

$A$ is required to be unit $l_2$-norm (or bounded norm) in order to avoid the trivial solutions that are due to the ambiguity of the linear reconstruction.

Step 2: Construct and solve an $l_1$-norm minimization problem,

$$\arg \min_x \|x\|_1, s.t., Ax = y \qquad (2)$$

where $x$ is called as spare representation coefficients of $y$.

Eq. (2) can be usually approximate by an $l_1$-norm minimization problem,

$$\arg \min_x \|x\|_1, s.t., \|Ax-y\|_2 \leq \varepsilon \qquad (3)$$

where $\varepsilon$ is the threshold of the residue.

Eq.(3) can be generalized as a constrained least square problem,

$$\arg \min_x \|Ax-y\|_2^2 + \lambda \|x\|_1 \qquad (4)$$

where $\lambda>0$ is a scalar regularization parameter which balances the tradeoff between the sparsity of the solution and the reconstruction error.

Eq.(4) is a constrained LASSO problem, its detail solution is found in Ref. [27].

Step 3: Compute residue $r_i(y) = \|y - A\delta_i\|_2, i=1,2,...,C$, where $\delta_i: R^n \to R^n$ is the characteristic function that selects the coefficients associated with the $i^{th}$ class;

Step 4: the class label of, $y$ $C(y)$, is identified as $C(y) = \arg \min_i r_i(y)$.

## 2.3 WSRC

WSRC integrates both sparsity and locality structure of the data to further improve the classification performance of SRC. It aims to impose larger weight to the training samples that are 'farer' from the test sample. Different from

SRC, WSRC solves a weighted $l_1$-norm minimization problem,

$$\arg\min_{x} \|Ax\text{-}y\|_2^2 + \lambda \|Wx\|_1 \tag{5}$$

where $W$ is a diagonal weighted matrix, and its diagonal elements are $w_1, w_2, ..., w_n$.

Eq.(5) makes sure that the coding coefficients of WSRC tend to be not only sparse but also local in linear representation [13], which can represent the test sample more robustly.

## 2.4 LSRC

Though a lot of instances have been reported that WSRC performs better than SRC in various classification problems, WSRC forms the dictionary by using all the training samples, thus the size of the generated dictionary may be large, which will make adverse effect to solving the $l_1$-norm minimization problem. To overcome this drawback, a local sparse representation based classification (LSRC) is proposed to perform sparse decomposition in a local manner. In LSRC, K-NN criterion is exploited to find the nearest $k$ neighbors for the test samples, and the selected samples are utilized to construct the over-complete dictionary. Different from SRC, LSRC solves a weighted $l_1$ minimization problem,

$$\arg\min_{x} \|A_{N(y)}x\text{-}y\|_2^2 + \lambda \|x\|_1 \tag{6}$$

where $A_{N(y)} \in R^{m \times k}$ stands for data matrix which consists of the $k$ nearest neighbors of $y$.

Compared with the original SRC and WSRC, although the computational cost of LSRC will be saved remarkably when $k \ll n$, LSRC does not assign different weight to the different training samples.

## 3. Two-stage LSCL

From the above analysis, it is found that each of LMC, WSRC and LSRC has its advantages and disadvantages. To overcome the difficult problem of plant recognition on the large-scale leaf image database, a two-stage LSCL leaf recognition method is proposed in the section. It is a sparse decomposition problem in a local manner to obtain an approximate solution. Compared with WSRC and LSRC, LSCL solves a weighted $l_1$-norm constrain least square problem in the candidate local neighborhoods of each test sample, instead of solving the same problem for all the training samples. Suppose there are a test sample $y \in R^m$ and $n$ training samples from $C$ classes $\{X_1, X_2, ..., X_C\}$, and $n_i$ is the sample number of $i$th class $X_i$, $x_{ij}$ $(j=1,2,...,n_i)$ is $j$th sample of the $i$th class $X_i (i=1,2,...,C)$. Each sample is assumed to be a one-dimensional column vector. The proposed method is described in detail as follows.

## 3.1 First stage of LSCL

(1) Calculate the Euclidean distance $d(y, x_{ij})$ between $y$ and $x_{ij}$, and select $k$ nearest neighbors of $y$ from $X_i$ with the first $k$ smallest distances $d(y, x_{ij})$, the selected neighbor subset noted as $N(X_i) = \{x_{i1}^k, x_{i2}^k, ..., x_{ik}^k\}$, $i = 1, 2, ..., C$.

(2) Calculate the average of $N(X_i)$,

$$\overline{X_i} = \frac{1}{k}\sum_{j=1}^{k} x_{ij}^k \tag{7}$$

(3) Calculate the Euclidean distance $d(y, \overline{X_i})$ between $y$ and $\overline{X_i}$.

(4) From $C$ neighbor subsets, select $S$ neighbor subsets with the first $S$ smallest distances $d(y, \overline{X_i})(i=1,2,...,C)$ as the candidate subsets for the test sample, in simple terms, denoted as $N(X_1), N(X_2), ..., N(X_S)$.

The training samples from $N(X_1), N(X_2), ..., N(X_S)$ are reserved as the candidate training samples for the test sample, and the other training samples are eliminated from the training set.

## 3.2 Second step of LSCL

From the first stage, it is noted that there are $p = k \cdot S$ training samples from all the candidate subsets $N(X_1), N(X_2), ..., N(X_S)$. For simplify, we just as well express the $j$th training sample of $N(X_i)$ is $x_{ij}(i=1,2,...,S; j=1,2,...,k)$. The second stage first represents the test sample as a linear combination of all the training samples of $N(X_1), N(X_2), ..., N(X_S)$, and then exploits this linear combination to classify the test sample.

From the first stage, we have obtained the Euclidean distance $d(y, x_{ij})$ between $y$ and each candidate sample $x_{ij}(i=1,2,...,S; j=1,2,...,k)$. By $d(y, x_{ij})$, a new local WSRC is proposed to solve the same weighted $l_1$-norm minimization problem as Eq.(5),

$$\arg\min_{x} \|Ax - y\|_2^2 + \lambda \|W'x\|_1 \tag{8}$$

where $A \in R^{m \times p}$ is the dictionary constructed by $p = k \cdot S$ training samples of $N(X_1), N(X_2), ..., N(X_S)$, $\text{diag}(W') = [d(y, x_{11}), d(y, x_{12}), ..., d(y, x_{1k}), ..., d(y, x_{Sp})]^T$ is the weighted diagonal matrix, $d(y, x_{ij})$ is the Euclidean distance between $y$ and $x_{ij}$.

In Eq.(8), the weighted matrix is a locality adaptor to penalize the distance between $y$ and $x_{ij}$. In the above SRC, WSRC, LSRC and LSCL, the $l_1$−norm constraint least square minimization problem is solved by the approach proposed in [28], which is a specialized interior-point method for solving the large scale problem. The solution of Eq.(8) can be expressed as

$$\tilde{x} = \left(A^T A + \lambda W'\right)^{-1} A^T y \tag{9}$$

From Eq.(9), $A\tilde{x}$ is expressed as the sparse representation of the test sample. In representing the test sample, the sum of the contribution of the $i$th candidate neighbor subset is calculated by

$$y_i' = \tilde{x}_{i1} \cdot x_{i1} + \tilde{x}_{i2} \cdot x_{i2} + ... + \tilde{x}_{ik} \cdot x_{ik} \qquad (10)$$

where $\tilde{x}_{ij}$ is the $j$th sparse coefficient corresponding to the $i$th candidate nearest neighbor subset.

Then we calculate the residue of the $i$th candidate neighbor subset corresponding to test sample $y$,

$$r_i(y) = \|y - y_i'\|_2, i = 1, 2, ..., S \qquad (11)$$

In Eq.(11), for the $i$th class ($i=1,2,...,S$), a smaller $r_i(y)$ averages the greater contribution to representing $y$. Thus, $y$ is finally classified into the class that produces the smallest residue.

## 3.3 Summary of two-stage LSCL

From the above analysis, the main steps of the proposed method are summarized as follows.

Suppose $n$ training samples from $C$ different classes, a test sample $y$, the number $k$ of the nearest neighbors of $y$, the number $S$ of the candidate neighbor subsets.

Step 1. Compute the Euclidean distance between the test sample $y$ and every training sample $x_{ij}$, respectively.

Step 2. Through K-NN rules, find $k$ nearest neighbors from each training class as the neighbor subset for $y$, calculate the neighbor average of the neighbor subset of each class, and calculate the distance between $y$ and the neighbor average.

Step 3. Determine $S$ neighbor subsets with the first $S$ smallest distances, as the candidate neighbor subsets for $y$.

Step 4. Construct the dictionary by all training samples of the $S$ candidate neighbor subsets and then construct the weighted $l_1$-norm minimization optimization problem as Eq.(8).

Step 5. Solve Eq.(8) and obtain the sparse coefficients.

Step 6. For each candidate neighbor subset, compute the residue between $y$ and its estimation $y_i'$ by Eq.(11).

Step 7. Identify the class label that has the minimum ultimate residue and classify $y$ into this class.

## 3.4 Rationale and interpretation of LSCL

In practical, some between-species leaves are very different from the other leaves, as shown in Fig.1A. They can be easily classified by the Euclidean distances between the leaf digital image matrices. However, some between-species leaves are very similar to each other, as shown in Fig.1B. They cannot be easily classified by some simple classification methods. In Figs.1A and B, suppose the first leaf is the test sample, while other seven leaves are training samples. It is difficult to identify the label of the test leaf by the simple classification method, because the test leaf is very similar to Nos. 4,5,6 and 7 in Fig.1B. However, it is sure that the test sample is not Nos.1, 2 and 3. So, we can naturally firstly exclude these three leaves. This exclusion method example is the purpose of the first stage of LSCL. From Fig.1C, it is found that there is large difference between the leaves of the same species. Therefore, in plant recognition, an optimal scheme is to select some training samples that are relatively similar to the test sample as the candidate training samples, such as Nos. 2 and 9 in Fig.1C are similar to the test sample in Fig.1C, instead of considering all training samples. The average neighbor distance is used to coarsely recognize the test sample. The average neighbor distance as dissimilarity is more effective and robust than the original distance between the test and each training leaf, especially in the case of existing noise and outliers.

From the above analysis, in the first stage of LSCL, it is reasonable to assume that the leaf close to the test sample has great effect, on the contrary, if a leaf is far enough from the test sample it will have little effect and even have side-effect on the classification decision of the test sample. These leaves should be discarded firstly, and then the later plant recognition task will be clear and simple. In the same way, we can use the similarity between the test sample and the average of its nearest neighbors to select some neighbor subsets as the candidate training subsets of the test sample. If we do so, we can eliminate the side-effect on the classification decision of the neighbor subset that is far from the test sample. Usually, for the classification problem, the more the classes, the lower the classification accuracy, so the first stage is very useful.

In the second stage of LSCL, there are $S$ nearest neighbor subsets as candidate class labels of the test sample, thus it is indeed faced with a problem simpler than the original classification problem, because $S < C$ and $k \cdot S \ll n$, i.e., few training samples are reserved to match the test sample. Thus, the computational cost is mostly reduced and the recognition rate will be improved greatly. We analyze the computational cost of LSCL in theory as follows.

There are $n$ samples from $C$ classes, and every sample is an $m \times 1$ column vector, the first stage need to calculate the Euclidean distance, select $k$ nearest neighbors from each class, and calculate the average of the $k$ nearest neighbors, then the computational cost is about $O(mn^2/k)$. In second stage, there are $p = k \cdot S$ training samples to construct the dictionary $A$, the cost of $B = A^T A$ is $O(mp^2)$, the cost of $D = (B + \lambda W')^{-1}$ is $O(p^3)$, and the cost of $DA^T y$ is $O(mp^2) + O(mp)$. The second stage has computational cost of $O(p^3) + O(mp^2) + O(mp)$. The computational cost of LSCL is $O(mn^2/k) + O(p^3) + O(mp^2) + O(mp)$ in total. The computational cost of the classical SRC algorithm is $O(n^3) + O(mn^2) + O(mn)$ [8,9]. Compared with SRC, it is found that the computational cost of LSCL will be saved remarkably when $k \cdot S < n$.

## 4. Experiments and result analysis

In this section, the proposed method is validated on a plant species leaf database and compared with the state-of-the-art methods.

**4.1 Leaf image data and experiment preparation**

To validate the proposed method, we apply it to the leaf classification task using the ICL dataset. All leaf images of the dataset were collected at the Botanical Garden of Hefei, Anhui Province of China by Intelligent Computing Laboratory (ICL), Chinese Academy of Sciences. The ICL dataset contains 6000 plant leaf images from 200 species, in which each class has 30 leaf images. Some examples are shown in Fig.2. In the database, some leaves could be distinguished easily, such as the first 6 leaves in Fig.2A, while some leaves could be distinguished difficultly, such as the last 6 leaves in Fig.2A. We verify the proposed method by two situations, (1) two-fold cross validation, i.e., 15 leaf images of each class are randomly selected for training, and the rest 15 samples are used for testing; (2) leave-one-out cross validation, i.e., one of each class are randomly selected for testing and the rest 29 leaf images per class are used for training.

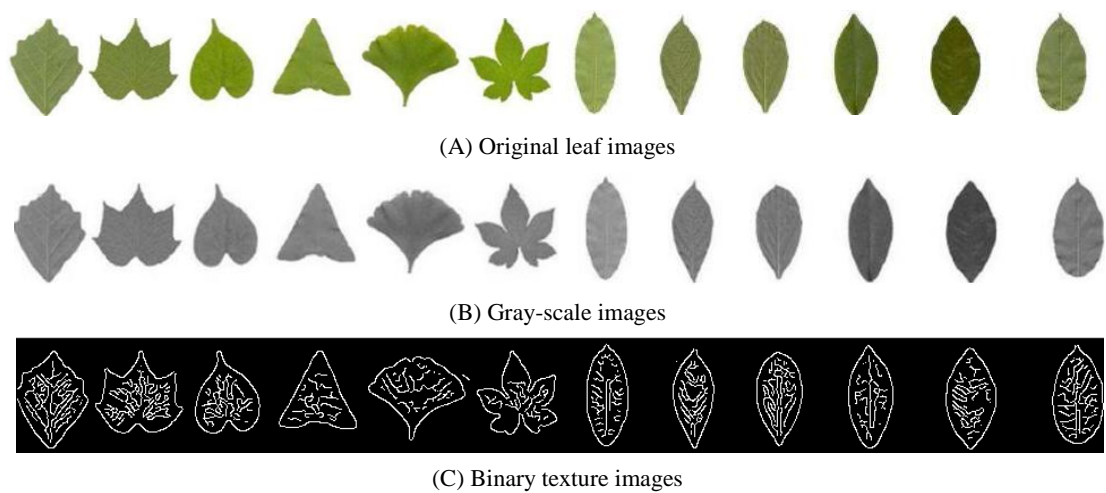

(A) Original leaf images

(B) Gray-scale images

(C) Binary texture images

Fig.2 Samples of different species from ICL database

All these experiments are repeated many times; consequently the average recognition rates are reported. To show the performance of the proposed algorithm, we compare our recognition performance with five existing plant recognition methods for several reasons, three feature extraction based leaf recognition methods, i.e., Multiscale Distance Matrix (MDM) [20], Shape Features and Colour Histogram (SFCH) [21], texture and shape features (TSF) [22], and two sparse representation based leaf recognition methods, i.e., learning sparse representation (LSR)[23] and sparse representation (SR)[24]. Firstly, we conduct the experiments on the same database as in MDM. Secondly, SFCH and TSF are two current plant recognition methods. Thirdly, SRC has been rarely applied to leaf based plant recognition. All experiments are carried out using MATLAB 7.0 software on an Intel Xeon X3430 PC with 2 GB RAM. In addition, the MATLAB functions 'K-NNclassify' and 'edge' are used to find the nearest neighbors and extract the texture images, respectively, and the SR coefficients and residues is solved by the function of 'SRC' in SR Toolbox (http://sites.google.com/site/sparsereptool).

### 4.2 Pre-processing

Each leaf image is firstly converted to gray-scale by

$$Gray = 0.2989*R + 0.5870*G + 0.1140*B \qquad (12)$$

where Gray is the grayscale value, R is the red component, G is the green component and B is the blue component, as shown in Fig.2B.

Canny edge detection algorithm is one of the most strictly defined methods to provide good and reliable edge detection [29], which can be implemented by MATLAB function 'edge'. We extract the texture image from each leaf through 'edge' function, as shown in Fig.2C. Each texture image is cropped and normalized to the same size of 32×32, and concatenated to a feature vector of size 32×32=1024. There is no further extra feature extraction step.

### 4.4 Classification process

For each test sample $y$, we compute the Euclidean distance between $y$ and every training sample $x_{ij}$, find the $k$ nearest neighbors as neighbor subset for $y$ from each training class by using K-NN criterion, calculate the neighbor average of the $k$ nearest neighbors of each class, and calculate the distance between $y$ and each neighbor average, finally determine $S$ neighbor subsets with the first $S$ smallest distances, as the candidate neighbor subsets for the test sample. Then there are $k \cdot S$ training samples from the $S$ candidate nearest neighbor subsets.

Construct the dictionary by $k \cdot S$ training samples of the $S$ candidate neighbor subsets and then construct the weighted $l_1$-norm minimization optimization problem Eq.(8), which is solved by $l_1$-regularized least squares [11,13,27]. Then the sparse coefficients are obtained, and finally the test sample is classified into the class that produces the smallest residue.

### 4.5 Parameter setting

In nature, most of between-species leaves are very different from each other. Given a test sample, it is reasonable to think that more than half of leaves from database are far from the test sample, so the number of the candidate subsets $S$ can be set about half of the number of all training classes. We set $S = \lceil 200/3 \rceil = 70$. That is to say, 200-70=130 classes are excluded directly by the neighbour average, thus 70 candidate neighbor subsets are reserved with 10 training samples in each subset, i.e., in total 10×70=700 are reserved as candidate training samples. The parameter $\lambda$ is empirically set to 0.01 to achieve the best results. We prepare experiments to investigate the effect of the parameter $k$. In the experiments, the nearest neighbour size $k$ is an important parameter. Some experiments are conducted to evaluate the effect on the different $k$. For each $k$ in range [5,25], by two-fold cross validation, we repeated 20 times and the averaged results are shown in Fig. 3. The results validated that the proposed method can achieve the better recognition rate when $k$ is 13. In the following experiments, $k$ is simply set 13.

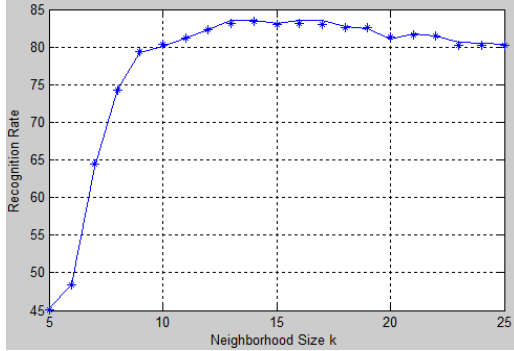

Fig.3. Experiments by leave-one-out cross validation scheme with different $k$

### 4.6 Experimental results

In two-fold cross validation scheme, we conduct 50 independent experiments on randomly selecting 15 samples from each class as training subset, while the remaining 15 samples as test subset. In leave-one-out cross validation scheme, we conduct 30 independent experiments on each sample as once test sample. The recognition accuracy per class is reported over all independent experiments, where the accuracy per class is the percentage of the successful recognitions of each class relative to the number of leaves from the corresponding class. In all experiments, we observe that the classification accuracy rates in the first stage are 100% on ICL database. The reason is that most between-species leaves are different from each other, which can be coarsely classified by the simple K-NN based clustering and elimination algorithms, such as LMC used in the first stage of the proposed method. Table 1 shows the average correct recognition results. To show the effectiveness of the proposed method, Table 1 also shows the recognition results by other 5 methods.

Table 1 Average recognition rates, standard deviation (percent) and running time (seconds) of MDM, SFCH, TSF, LSR, SR and the proposed method by half-fold cross validation scheme

| Method | MDM | SFCH | TSF | LSR | SR | Our method |
| --- | --- | --- | --- | --- | --- | --- |
| Recognition results | 0.7643 ±0.230 | 0.7496 ±0.264 | 0.7537 ±0.257 | 0.7482 ±0.213 | 0.7576 ±0.236 | 0.8139 ±0.204 |
| Running time($s$) | 108 | 128 | 144 | 107 | 116 | 86 |

Table 2 Average recognition rates, standard deviation (percent) and running time (seconds) of MDM, SFCH, TSF, LSR, SR and the proposed method by leave-one-out cross validation scheme

| Method | MDM | SFCH | TSF | LSR | SR | Our method |
| --- | --- | --- | --- | --- | --- | --- |

| Recognition results | 0.8203 ±0.205 | 0.8065 ±0.214 | 0.8187 ±0.196 | 0.8125 ±0.125 | 0.8194 ±0.176 | 0.8412 ±0.168 |
|---|---|---|---|---|---|---|
| Running time(*s*) | 89 | 106 | 112 | 84 | 96 | 74 |

## 4.7 Result analysis

From Tables 1 and 2, it is found that the proposed method outperforms the other three leaf feature extraction based plant recognition methods and two SR based plant recognition algorithms, and achieves the highest average recognition results and lowest running time. The possible explanation, to some extent, is that (1) other 5 methods, in fact, are feature extraction based plant recognition methods. They firstly extract the different features and then use different classifiers to classify the leaf images. But, from Fig.1C, it is found that the leaves of the same species are very different from each other, so it is difficult to effectively extract the optimal features from each leaf image. Moreover, it take much time to extract the features from each leaf image, so the computational cost of the comparative 5 methods are larger than the proposed method; (2) the proposed method fully utilize the distance between the test sample and the average of its nearest neighbors to select the candidate neighbor subsets, which can eliminate the side-effect on the classification decision, thus the sparse representation result is sparser than that of the classical SR algorithm. Moreover, compared with LSR[23] and SR[24], using the relatively few samples of the candidate neighbor subsets to represent the $l_1$-minimization optimal problem, much time is saved; (3) especially, the main reason is that the proposed method is robust because selecting $k$ nearest neighbors and the weighted SRC algorithm can restrain the influence of outliers and noise.

In the existing SRC, WSRC and LSRC methods, SRC and WSRC represent the test sample by all the training samples, and calculates SR coefficients by solving Eqs.(4) and (5), respectively, while LSRC by solving (6). If we apply directly them to leaf recognition, there are more non-zero reconstruction coefficients, because there are a lot of similar leaves from different species, as shown in Fig.1B and Fig.2A. As a result, the computation cost to find the 'sparse' solution tends to increase greatly. Although they have been successfully applied to face recognition, they are not optimal methods for the plant recognition on the large leaf image database, because (1) too many intra-class different and inter-class similar leaf images in the training set make the SR of the test sample not to be sparse. So the classical SRC and WSRC is not an optimal choice for the leaf based plant recognition on the large leaf image database; (2) when the number of the training samples increases, the computational cost to solve Eqs.(4) and (5) will increase quickly. In this sense, we firstly eliminate a lot of training samples by selecting $k$ nearest neighbors and the candidate neighbor subsets. Two-stage plant recognition scheme can improve the performance of the plant recognition method and reduce the computation cost. Although LSRC can reduce the computational complexity implemented in the local neighbors of each test samples, it does not consider the prior similar relationship between the test sample and the individual training sample, which may enhance the classification performance of SRC in principle. As a whole, the recognition result and computational time of the proposed method is acceptable for plant recognition on a large-scale database.

## 5. Conclusion

Plant species recognition is important and difficult. This paper divides the procedure of plant species recognition into two stages. The first stage directly exploits the Euclidean distance to decide the nearest neighbor subsets that are ''close'' to the test sample and exclude the other training samples. The second stage represents the test sample as a linear combination of all the training samples from the candidate neighbor subsets, and exploits the residue to classify the test sample. Because our method uses only a few subsets of all the training samples to represent the test sample, the classification problem becomes simpler and the time cost reduces greatly. In other words, the original classification problem needs to assign the test sample into one of all the original classes, whereas the classification problem in the second stage of the proposed method just needs to assign the test sample into one of few nearest neighbor subsets.

Usually, to assign the test sample into one of few classes is simpler and will obtain higher accuracy. Experimental results illustrate the good performance of our method. But in general, sparse representation needs an over complete dictionary. In the LSRC and the proposed method, the condition may be unsatisfied. It will be explained in theory in the future work. In experiments we also observed that the size of nearest neighbors to correctly classify a test sample is different. In fact in the K-NN classifier, some test samples need a small nearest neighbors to yield the correct results; meanwhile some other test samples need a larger nearest neighbors to correctly classify. Therefore it would be interesting to design an adaptive scheme to adjust the nearest neighbor size for different classes in the proposed method. This issue will be further investigated in future.

### Acknowledgments

This work is partially supported by the China National Natural Science Foundation under grant Nos. 61473237 and 61309008. It is also supported by the Shaanxi Natural Science Foundation Research Project under grant No. 2014JM2-6096.The authors would like to thank all the editors and anonymous reviewers for their constructive advices.

## References

[1] Mitani Y., Hamamoto Y. Classifier design based on the use of nearest neighbor samples. *Proc. 15th Int. Conf. Pattern Recognition*, Barcelona, Spain, 2000:769–772.

[2] Samsudin N.A., Bradley A.P. Nearest neighbor group-based classification. Pattern Recognition, 2010,43, 3458–3467.

[3] Yang J., Zhang L., Yang J.Y., Zhang David. From classifiers to discriminators: a nearest neighbor rule induced discriminant analysis. Pattern Recogn., 2011,44, 1387–1402.

[4] Chai J., Liu H., Chen B., et al. Large margin nearest local mean classifier. Signal Process.,2010,90, 236–248.

[5] Mitani Y., Hamamoto Y. A local mean-based nonparametric classifier. Pattern Recognition Letters,2006, 27(10), 1151-1159.

[6] Xiaoqin Zhang, Feng Liu. A New Local Mean-based Nonparametric Classification Method. 2nd International Conference on Electrical, Computer Engineering and Electronics (ICECEE), 2015:1560-1564

[7] Marti nez, A.M, Kak A.C. PCA versus LDA. IEEE Trans. Patt. Anal. Mach. Int. 2001, 23, 228-233.

[8] J. Wright, A.Y. Yang, A. Ganesh, et al. Robust face recognition via sparse representation, IEEE Transactions on Pattern Analysis and Machine Intelligence 31 (2009) 210–227.

[9] J. Wright, Y. Ma, J. Mairal, et al. Sparse representation for computer vision and pattern recognition, Proceedings of the IEEE 98 (2010) 1031–1044.

[10] W.Deng, J.Hu, J.Guo. Extended SRC: under sampled face recognition via intraclass variant dictionary. IEEE Trans. PatternAnal.Mach.Intell.,34(9) (2012)1864–1870.

[11] J. Yang, L. Zhang, Y. Xu, et al. Beyond sparsity: the role of L1-optimizerin pattern classification, Pattern Recognition, 2012,45(3):1104–1118.

[12] Song Guo, Qiuqi Ruan, Zhenjiang Miao. Similarity Weighted Sparse Representation for Classification. International Conference on Pattern Recognition (ICPR),2012:1241-1244

[13] C.Y. Lu, H. Min, J. Gui, et al. Face recognition via weighted sparse representation. Journal of Visual Communication and Image Representation,2013,24(2):111-116.

[14] Chun-Guang Li, Jun Guo, Hong-Gang Zhang. Local Sparse Representation based Classification. International Conference on Pattern Recognition,2010:649-653.

[15] Zhang S. W, Lei Y K, Wu, Y. H. Semi-supervised locally discriminant projection for classification and recognition. Knowl.-Based Syst., 2011, 24, (2):341-346.

[16] Zhang S W, Lei Y K. Modified locally linear discriminant embedding for plant leaf recognition. Neurocomputing, 2011,74, (14-15):2284-2290.

[17] Zhao C, Chan S F, Cham W K, et al. Plant identification using leaf shapes - A pattern counting approach. Pattern Recognition,2015,48 (10),3203-3215.

[18] Trishen Munisami, Mahess Ramsurn, Somveer Kishnah, et al. Plant Leaf Recognition Using Shape Features and Colour Histogram with K-nearest Neighbour Classifiers. Procedia Computer Science,2015, 58:740-747

[19] Nisar Ahmed, Usman Ghani Khan, Shahzad Asif. An automatic leaf based plant identification system. Sci.Int.(Lahore),28(1),427-430,2016.

[20] Hu R, Jia W, Ling H, et al. Multiscale Distance Matrix for Fast Plant Leaf Recognition. IEEE Transactions on Image Processing A Publication of the IEEE Signal Processing Society, 2012, 21(11):4667-4672.

[21] Munisami T, Ramsurn M, Kishnah S, et al. Plant Leaf Recognition Using Shape Features and Colour Histogram with K-nearest Neighbour Classifiers. Procedia Computer Science, 2015, 58:740-747.

[22] Chaki J, Parekh R, Bhattacharya S. Plant leaf recognition using texture and shape features with neural classifiers. Pattern Recognition Letters, 2015, 58(C):61-68.

[23] Hsiao J K, Kang L W, Chang C L, et al. Learning sparse representation for leaf image recognition. IEEE International Conference on Consumer Electronics. IEEE, 2014:209-210.

[24] Taisong Jin, Xueliang Hou, Pifan Li, et al. A Novel Method of Automatic Plant Species Identification Using Sparse Representation of Leaf Tooth Features. PLoS One. 2015, 10(10):1-20.

[25] Ran He, BaoGang Hu, Wei-Shi Zheng, et al. Two-stage Sparse Representation for Robust Recognition on Large-Scale Database. Proceedings of the Twenty-Fourth AAAI Conference on Artificial Intelligence (AAAI-10),2010:475-480

[26] Yong Xu, Qi Zhu, Zizhu Fan, et al. Using the idea of the sparse representation to perform coarse-to-fine face recognition. Information Sciences 238 (2013) 138–148.

[27] Elhamifar E, Vidal R. Sparse subspace clustering: algorithm, theory, and applications. IEEE T Pattern Anal Mach Intel.2013; 35(11): 2765–2781

[28] S.J. Kim, K. Koh, M. Lustig, et al. A method for large-scale $l_1$-regularized least squares. IEEE Journal on Selected Topics in Signal Processing, 2007,1(4):606–617.

[29] Canny J. A Computational Approach to Edge Detection. IEEE Trans. Pattern Analysis and Machine Intelligence, 1986, 8(6):679–698.